\documentclass[aps,pre,longbibliography,reprint,superscriptaddress]{revtex4-2}

\usepackage{mathrsfs}
\usepackage{amssymb}
\usepackage{amsmath}
\usepackage{amsfonts}
\usepackage{graphicx}
\usepackage{bm}
\usepackage{times,txfonts}
\usepackage{color}
\usepackage{physics}
\usepackage{braket}
\usepackage{lipsum}
\usepackage[colorlinks=true,linkcolor=blue,urlcolor=blue,citecolor=blue]{hyperref}
\usepackage{subfigure}
\usepackage[export]{adjustbox}

\begin{document}

\title{Thermodynamics of photoelectric devices}

\author{Samuel L. Jacob}
\email{samjac91@gmail.com}
\affiliation{School of Physics, Trinity College Dublin, College Green, Dublin 2, D02K8N4, Ireland}

\author{Artur M. Lacerda}
\email{machadoa@tcd.ie}
\affiliation{School of Physics, Trinity College Dublin, College Green, Dublin 2, D02K8N4, Ireland}

\author{Yonatan Dubi}
\email{jdubi@bgu.ac.il}
\affiliation{Department of Chemistry, Ben Gurion University of the Negev, Beer Sheva 8410501, Israel}

\author{John Goold}
\email{gooldj@tcd.ie}
\affiliation{School of Physics, Trinity College Dublin, College Green, Dublin 2, D02K8N4, Ireland}
\affiliation{Trinity Quantum Alliance, Unit 16, Trinity Technology and Enterprise Centre, Pearse Street, Dublin 2, D02YN67, Ireland}

\date{\today}

\begin{abstract}

We study the nonequilibrium steady state thermodynamics of a photodevice which can operate as a solar cell or a photoconductor, depending on the degree of asymmetry of the junction. The thermodynamic efficiency is captured by a single coefficient of performance. Using a minimal model based on a two-level system, we show that when the Coulomb interaction energy matches the transport gap of the junction, the photoconductor displays maximal response, performance and signal-to-noise ratio, while the same regime is always detrimental for the solar cell. Nevertheless, we find that the Coulomb interaction is beneficial for the solar cell performance if it lies below the transport gap. Our work sheds important light on design principles for thermodynamically efficient photodevices in the presence of Coulomb interactions.

\end{abstract}

\maketitle

\section{Introduction}

An understanding of the principles of energy conversion at small scales is crucial for the design of efficient nanodevices \cite{Benenti2017,Vinjanampathy2016,Binder2018,diVentra2008,Nazarov2009,Auffeves2022,Metzler2023}. In particular, photoelectric devices make use of light to generate an electrical current, holding great promise as highly controllable, clean and efficient energy converters -- with solar cells being a prime example \cite{Wuerfel2005,Nayak2019,AlEzzi2022,Solak2023}. From a thermodynamic point of view, solar cells can be viewed as continuous heat engines where the temperature gradient between the hot photon bath (the sun) and the cold bath (the cell) is used to generate an electric current. The efficiency of an ideal solar cell was studied in the seminal work by Queisser   and Shockley \cite{Shockley1961}, while subsequent works have identified fundamental losses \cite{Hirst2010,Markvart2022} and studied the efficiency at maximum power in nanosized solar cells \cite{Rutten2009,Ajisaka2015,Sarkar2020}. In solar cells composed of organic materials, inorganic quantum dots or carbon nanotubes, it is known that their high exciton binding energy -- the energy required to dissociate an excited electron-hole pair into free carriers -- is detrimental to their efficiency 
\cite{Giebink2011,Nayak2019,Zhu2021}. 


A less appreciated example of a photodevice is a photoconductor, where an imposed electric current changes in response to light. These are particularly relevant in molecular junctions, where the sensitivity of molecules to light can be used as a characterization and control tool to realize, e.g. photoswitches \cite{Tao2006,Cuevas2017,Dubi2011,Galperin2012,Aradhya2013,Xiang2016,Wang2017,Tang2022}. It is known that the conductance of the junction depends on factors such as the degree of junction asymmetry and exciton binding effects \cite{Cuevas2017,Dubi2017,Tang2022}. Junction asymmetry can arise due to different composition of the electrodes or due to the molecule itself being asymmetric. In both cases, the asymmetry alters the transport properties of the junction, which can be used e.g. for rectification \cite{VanDyck2015,Thong2017}. Regarding exciton binding effects, it was shown in Ref.~\cite{Dubi2017} that photoconductance in a symmetric molecular junction arises due to exciton binding. This counter-intuitive result indicates that exciton binding is not always detrimental for the operation of photoelectric devices. Nevertheless, a thermodynamic study of photoconducting devices is lacking.

The framework of stochastic thermodynamics, both classical \cite{Seifert2008,Seifert2012,Broeck2013,denBroeck2015} and quantum \cite{Kosloff2013,Vinjanampathy2016,Binder2018}, is the natural setting to study the thermodynamic efficiency of such small energy converters. Its central object is the entropy production (see \cite{Landi2021} for a review), quantifying the degree of irreversibility of physical processes within the device and constraining the efficiency of any heat engine below Carnot efficiency. In both classical and quantum devices operating at steady state, entropy production has also been shown to constraint current fluctuations through inequalities dubbed thermodynamic uncertainty relations \cite{Barato2015,Pietzonka2016,Pietzonka2018,Timpanaro2019,Horowitz2019,Guarnieri2019,Hasegawa2021,Kamijima2023,VanVu2023}. While there is, at present, a vast literature on the efficiency of heat engines \cite{Benenti2017,Myers2022} -- the solar cell being a particular case \cite{Rutten2009,Ajisaka2015,Sarkar2020} -- there is comparatively little focus on the thermodynamics of photoconducting devices. Ideally, one could use a microscopic model for a photodevice which captures its different operation regimes, i.e. solar cell and photoconductor regimes, in order to elucidate the underlying thermodynamics in a unified fashion.

In this work, we study the thermodynamics of a photodevice which operates as a photoconductor or solar cell, depending on the degree of asymmetry of the nanojunction. The steady state performance of the device is captured by a single coefficient of performance given by Eq.~\eqref{performance}. Our model incorporates the effect of Coulomb interaction within the junction (representing the exciton binding energy) which we show is critical for its performance: when the interaction energy matches the transport gap of the junction, a photoconductor reaches maximal performance and signal-to-noise ratio, while this regime is always detrimental for a solar cell. Nevertheless, we find that Coulomb interaction can be beneficial for the solar cell if it lies below the transport gap.

\begin{figure*}[t!]
        \centering
        \includegraphics[width=0.45\textwidth]{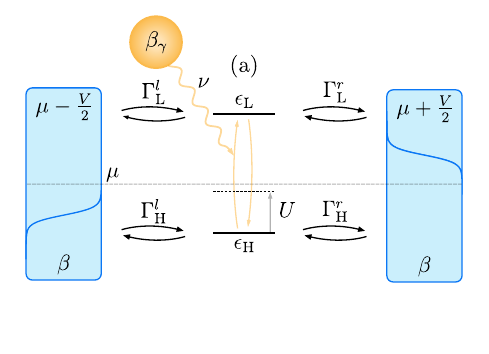}\hspace{0.9cm}
        \includegraphics[width=0.45\textwidth]{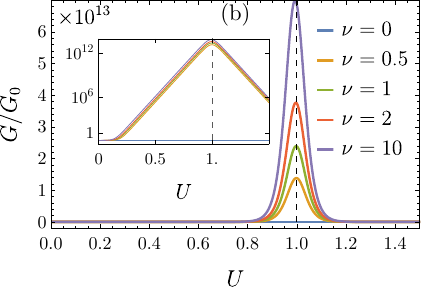}\hspace{0.5cm}
    \caption{(a) Scheme of the photodevice. The rates to the fermionic baths are determined by the quantum master equation in Eq.~\eqref{masterequation}. The Coulomb interaction energy $U$ causes a shift of the HOMO level towards the chemical potential $\mu$ and is responsible for photoconductance. (b) Conductance $G \equiv dJ/dV |_{V=0}$ as a function of interaction energy $U$ for different values of the photon rate $\nu$. The conductance is divided by $G_0$ representing the conductance for $U=0$ and $\nu=0$. The inset shows the same plot, but in semi-log scale. We set the rates $\Gamma^{l}_{\mathrm{H}} = \Gamma^{r}_{\mathrm{L}} \equiv \Gamma$ and $\Gamma^{l}_{\mathrm{L}} = \Gamma^{r}_{\mathrm{H}} = z~\Gamma$. Since all the rates in the quantum master equation appear as multiplicative constant, for numerical purposes only the ratio $\nu/\Gamma$ is relevant; we thus set $\Gamma = 1$ for simplicity and $z=1$ (symmetric device). The inverse temperatures are $\beta = 39.2$ and $\beta_{\gamma} = 2$ (corresponding to room temperature $295$K and sun temperature $5780$K when measured in electron-volt), $\mu = 0$, $\epsilon_{\mathrm{H}} = -1$ and $\epsilon_{\mathrm{L}} = 2$ (corresponding to an optical gap in electron-volts).}
    \label{fig1}
\end{figure*}

\section{Setup}

\subsection{Photodevice model \label{sec:photodevice}}

We consider a minimal model for a photoelectric device described by the total Hamiltonian $H = H_S + \sum_{\alpha} (H_{\alpha} + V_{S\alpha})$. The Hamiltonian $H_S = \sum_{i = \mathrm{H}, \mathrm{L}} \epsilon_i c^{\dagger}_i c_i + U c^{\dagger}_{\mathrm{H}} c_{\mathrm{H}} c^{\dagger}_{\mathrm{L}} c_{\mathrm{L}}$ describes a nanosystem with two electronic levels, such as a molecule, where $\{ c_i \}_{i=\mathrm{H,L}}$ are fermionic annihilation operators acting on the highest occupied molecular orbital (HOMO, $i=\mathrm{H}$) and lowest unoccupied molecular orbital (LUMO, $i=\mathrm{L}$) with $\epsilon_{\mathrm{L}} > \epsilon_{\mathrm{H}}$; the last term describes the Coulomb repulsion energy $U$ between two electrons present in the system (alternatively $-U$ can be interpreted as the exciton binding energy, which is the difference between the LUMO-HOMO gap and the transport gap \cite{Dubi2017}). The system interacts with three different baths $\alpha \in \{ l, r, \gamma \}$ described by the Hamiltonian $H_{\alpha} = \sum_{m_{\alpha}} \epsilon_{m_{\alpha}} d^{\dagger}_{m_{\alpha}} d_{m_{\alpha}}$ where $d_{m_{\alpha}}$ are either fermionic or bosonic annihilation operators and the sum is over all modes of the particular bath. The first two $\alpha = l, r $ are fermionic baths whose interaction with the system $V_{S\alpha} = \sum_{m_{\alpha}} \sum_{i = \mathrm{H,L}} (\lambda_{m_{\alpha},i}~c^{\dagger}_i d_{m_\alpha} + \mathrm{h.c.})$ allows for electron tunneling to and from HOMO and LUMO, thus defining the electronic junction. The bath $\alpha = \gamma$ is composed of photons and its interaction with the system $V_{S\gamma} = \sum_{m_{\gamma}} ( \lambda_{m_{\gamma}} c^{\dagger}_{\mathrm{L}} c_{\mathrm{H}}d_{m_{\gamma}} + \mathrm{h.c.})$ induces a transition between HOMO and LUMO. This defines the photodevice shown in the schematic of Fig.~\ref{fig1} (a).

\subsection{Quantum master equation}

We aim at a minimal model and as such all the baths are assumed to be ideal and the dynamics of the system will be described by a global quantum master equations of Lindblad form within the weak-coupling, Markovian and secular approximations \cite{CohenTannoudji1998,Breuer2007,Milburn2008,Rivas2012}. The effect of the fermionic baths $\alpha = l, r$ is then fully captured by the quantities $\{ \Gamma^{\alpha}_{\mathrm{H}}, \Gamma^{\alpha}_{\mathrm{L}}, f_{\alpha}(x)\}$ where $\Gamma^{\alpha}_{\mathrm{H}}, \Gamma^{\alpha}_{\mathrm{L}}$ are coupling rates to the HOMO and LUMO and $f_{\alpha}(x) = [\exp[\beta_{\alpha}(x-\mu_{\alpha})]+1]^{-1}$ is the Fermi-Dirac distribution at inverse temperature $\beta_{\alpha}$ and chemical potential $\mu_{\alpha}$. The photonic bath models thermal light and its effect is captured by $\{ \nu, n(x) \}$ where $\nu$ is the photon rate, namely, the rate at which photons are absorbed by an electron in the HOMO and pushed up to the LUMO and $n(x) = [\exp  (\beta_{\gamma} x) - 1]^{-1}$ is the Bose-Einstein distribution with $\beta_{\gamma}$ the inverse temperature of the radiation (see Appendix~\ref{app:qme} for the definition of these quantities). The quantum master equation in the Schr\"odinger picture is then given by 
\begin{align}
    \label{masterequation}
    \frac{d \rho_S(t)}{dt} = \mathcal{L}[\rho_S(t)] = -\frac{i}{\hbar}[H_S, \rho_S(t)] + \sum_{\alpha} \mathcal{D}_{\alpha}[\rho_S(t)] \; ,
\end{align}
where $\rho_S(t)$ denotes the density matrix of the system at time $t$ and we have ignored the effect of the Lamb-Shift Hamiltonian since it commutes with $H_S$. The last term describes dissipation induced by the baths $\mathcal{D}_{\alpha}[\cdot] = \sum_k \big[ L^{\alpha}_{k} \cdot {L^{\alpha}_k}^{\dagger} - 1/2 \{ {L^{\alpha}_k}^{\dagger} L^{\alpha}_k,\cdot  \} \big]$ where $L^{\alpha}_k$ are jump operators belonging to bath $\alpha$ (see Appendix~\ref{app:qme} for more details).

For a thermodynamic analysis, we focus on the average value of observables $\langle O \rangle_t \equiv \mathrm{Tr}[O \rho_S(t)]$ such as energy and particle number of the molecule $O = H_S, N_S$ where $N_S = c^{\dagger}_{\mathrm{H}} c_{\mathrm{H}} +c^{\dagger}_{\mathrm{L}} c_{\mathrm{L}}$, as well as its von Neumann entropy $S_S(t) = - \mathrm{Tr}[\rho_S(t) \log \rho_S(t)] \geq 0$. Their time derivative can be computed directly from Eq.~\eqref{masterequation}. For the energy and particle number, we obtain
\begin{align}
    \frac{d \langle H_S \rangle_t}{dt} & = \sum_{\alpha = l, r} [J_{Q}^{\alpha}(t) + \mu_{\alpha} J_N^{\alpha}(t)] + J_{Q}^{\gamma}(t) \label{energybalance} \; , \\
    \frac{d \langle N_S \rangle_t}{dt} & = \sum_{\alpha = l, r} J_{N}^{\alpha}(t) \label{particlebalance} \; .
\end{align}
In Eq.~\eqref{energybalance} we split the energy current coming from each bath into heat current 
\begin{align}
    \label{heatcurrent}
    J_{Q}^{\alpha}(t) = \mathrm{Tr}[(H_S - \mu_{\alpha} N_S) \mathcal{D}_{\alpha}[\rho_S(t)]]
\end{align}
and chemical work where 
\begin{align}
    \label{particlecurrent}
    J_{N}^{\alpha}(t) \equiv \mathrm{Tr}[N_S \mathcal{D}_{\alpha}[\rho_S(t)]] \; 
\end{align}
is the particle current. Note that for the photon bath $\mu_{\gamma} = 0$ so the energy current coincides with the heat current. Moreover, in Eq.~\eqref{particlebalance} there is no particle current induced by the photon bath since its interaction with the system does not change the particle number. Regarding the von Neumann entropy, it can be split into two contributions
\begin{align}
    \label{entropybalance}
    \frac{d S_S(t)}{dt} & = \dot{\Sigma}(t) + \sum_{\alpha=l,r,\gamma} \beta_{\alpha} J_{Q}^{\alpha}(t) \; .
\end{align}
The first term is the entropy production rate $\dot{\Sigma}(t) = -\sum_{\alpha} \mathrm{Tr}[\mathcal{D}_{\alpha}[\rho_S(t)] (\log \rho_S(t) - \log \omega_{\alpha})] \geq 0$ expressed as the time derivative of a relative entropy between $\rho_S(t)$ and the steady state imposed by each bath $\mathcal{D}_{\alpha}[\omega_{\alpha}] = 0$, while the second term describes the entropy current associated with the heat currents from each bath \cite{Spohn1978,Spohn1978a,Breuer2007}. For the fermionic baths $\alpha = l, r$ the steady state is the grand canonical ensemble $\omega_{\alpha} = Z_{\alpha}^{-1}e^{-\beta_{\alpha}(H_S - \mu_{\alpha}N_S)}$ where $Z_{\alpha} = \mathrm{Tr}[e^{-\beta_{\alpha}(H_S - \mu_{\alpha}N_S)}]$ is the partition function. For the photonic bath $\alpha = \gamma$, there is no unique steady state: since the photon bath acts only when a single electron is present in the system, any state with zero or two particles is left invariant under its action. However, we can choose the specific steady state $\omega_{\gamma} = Z_{\gamma}^{-1}e^{-\beta_{\gamma} H_S^{(1)}}$, with $Z_{\gamma} = \mathrm{Tr}[e^{-\beta_{\gamma}H^{(1)}_S}]$ and $H_S^{(1)} = \Pi_{1} H_S \Pi_{1}$ is the Hamiltonian restricted to the single particle sector by the projectors $\Pi_1$ onto this sector.

\begin{figure*}[t!]
    \centering
    \includegraphics[width=0.33\textwidth]{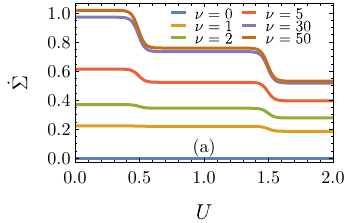}
    \includegraphics[width=0.33\textwidth]{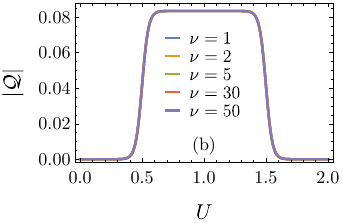}
    \includegraphics[width=0.33\textwidth]{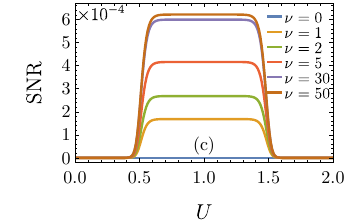}

    \vspace{0.2cm}
    
    \includegraphics[width=0.33\textwidth]{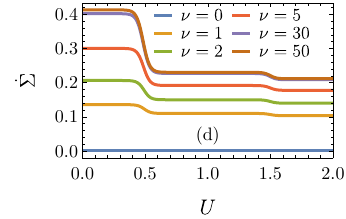}
    \includegraphics[width=0.33\textwidth]{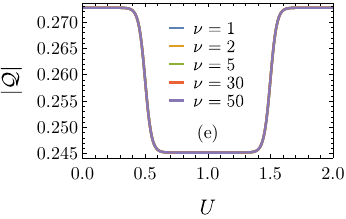}
    \includegraphics[width=0.33\textwidth]{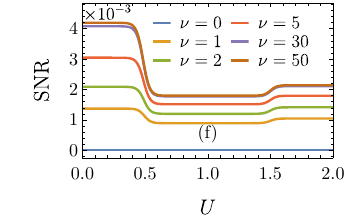}
    \caption{Steady state entropy production rate, absolute value of coefficient of performance and signal-to-noise ration (SNR) as a function of Coulomb interaction energy $U$ for different values of photon rate $\nu$. The first three panels (a)--(c) show a symmetric device ($z=1$) operating as a photoconductor $\mathcal{Q} < 0 $, while the last three panels (d)--(f) show an almost asymmetric device $(z=0.1$) operating as a solar cell $\mathcal{Q} > 0 $. Here $V=1$ and the remaining parameters are the same as in Fig.~\ref{fig1} (b).}
    \label{fig2}
\end{figure*}

\section{Results}

\subsection{Coefficient of performance}

We now restrict our analysis of the device to the steady state regime, where the left hand sides of Eqs.~\eqref{masterequation}, \eqref{energybalance}, \eqref{particlebalance} and \eqref{entropybalance} vanish. We henceforth drop the time dependence to refer to steady state quantities. From Eq.~\eqref{particlebalance} we see that the steady state is defined by a unique particle current $J \equiv J^{l}_{N} = -J^{r}_{N}$, where $J > 0$ when the current flows into the system from the fermionic bath $l$ (similarly for other currents from the other baths). Eqs.~\eqref{energybalance} and \eqref{entropybalance} can then be written $(\mu_{r} - \mu_{l})J = \sum_{\alpha} J^{\alpha}_{Q}$ and $\dot{\Sigma} = -\sum_{\alpha} \beta_{\alpha} J_Q^{\alpha} \geq 0$, expressing the first and second law of thermodynamics at steady state.

We now consider the fermionic baths to be held at the same temperature, lower than the photon bath $\beta_l = \beta_r = \beta > \beta_{\gamma}$, while their chemical potentials differ by a potential bias $V \geq 0$ such that $\mu_r = \mu + V/2 \geq \mu - V/2 = \mu_l$ where $\mu$ is the chemical potential without bias. Using the first law and second law, the entropy production rate can be written as
\begin{align}
    \label{entropyproduction}
    \dot{\Sigma} = \beta J_{Q}^{\gamma}(\eta_C - \mathcal{Q}) \geq 0 \; ,
\end{align}
where $\eta_C = 1 - \beta_{\gamma}/\beta$ is the Carnot efficiency and the coefficient of performance $\mathcal{Q}$ is defined by
\begin{align}
    \label{performance}
    \mathcal{Q} = VJ/J_{Q}^{\gamma} \; .
\end{align}
When $J^{\gamma}_{Q} > 0$ (as is the case in our model) the coefficient $\mathcal{Q}$ captures the steady state performance of our device across two different regimes. $\mathcal{Q} > 0$ describes the \textit{efficiency of a solar cell} -- characterized by a particle current $J > 0$ flowing against the potential bias -- and is bounded by Carnot according to Eq.~\eqref{entropyproduction}. $\mathcal{Q} < 0$ describes the \textit{performance of a photoconductor} -- characterized by a particle current $J < 0$ flowing with the potential bias -- and its absolute value is unbounded. In order to better elucidate the physics behind Eq.~\eqref{performance}, we use Eq.~\eqref{masterequation} to derive a master equation for the steady-state populations of the device. Thus we have $\dot{p}^{\alpha}_m = \sum_{n \neq m} j^{\alpha}_{mn}$, where $\dot{p}^{\alpha}_m$ is the rate of population change induced by bath $\alpha$ on level $m \in \{ 0, 1, 2, 3 \}$, corresponding in order to an empty device, one electron in HOMO, one electron in LUMO and doubly occupied; while $j^{\alpha}_{mn} = - j^{\alpha}_{nm}$ is the associated population current flowing from level $n$ to $m$ due to bath $\alpha$. We can then show that Eq.~\eqref{performance} can be recast as (see Appendix~\ref{app:performance})
\begin{align}
    \label{performancerate}
    \mathcal{Q} = \frac{V}{\epsilon_{\mathrm{L}} - \epsilon_{\mathrm{H}}} \Bigg[ 1 - \frac{j^{r}_{\mathrm{H}} -j^{l}_{\mathrm{L}}}{j^{\gamma}_{21}} \Bigg] \; .
\end{align}
In the last expression $j^{r}_{\mathrm{H}} \equiv j^{r}_{10} + j^{r}_{32}$ is the current entering HOMO from the fermionic bath $r$, $j^{l}_{\mathrm{L}} \equiv j^{l}_{20} + j^{l}_{31}$ is the current entering LUMO from the fermionic bath $l$ and $j^{\gamma}_{21}$ is the current from HOMO to LUMO induced by light. For solar cell operation $\mathcal{Q} > 0$ we must have $j^{\gamma}_{21} > j^{r}_{\mathrm{H}} - j^{l}_{\mathrm{L}}$. Thus, the solar cell regime is only possible if the rate of HOMO-LUMO excitation by light exceeds the current flowing \textit{with} the potential bias from the fermionic baths. In the ideal case of a totally asymmetric device, where the fermionic bath $r$ is disconnected from HOMO and $l$ is disconnected from LUMO, i.e. $\Gamma^{r}_{\mathrm{H}} = \Gamma^{l}_{\mathrm{L}} = 0$, these currents vanish $j^{r}_{\mathrm{H}} = j^{l}_{\mathrm{L}} = 0$. We thus conclude that a totally asymmetric device always operates as a solar cell with maximum efficiency $0 <\mathcal{Q} = V/(\epsilon_{\mathrm{L}} - \epsilon_{\mathrm{H}}) \leq \eta_C$. This efficiency represents the maximum efficiency attainable in the irreversible regime and has been obtained e.g. in Ref.~\cite{Rutten2009}. Our results show that such an efficiency is purely a consequence of the totally asymmetric character of the coupling, which renders the performance insensitive to the Coulomb interaction. For photoconductor operation $\mathcal{Q} < 0$ we must have the reverse condition $j^{r}_{\mathrm{H}} - j^{l}_{\mathrm{L}} > j^{\gamma}_{21}$ and the current flowing \textit{with} the potential bias from the fermionic baths has to be larger than the rate of HOMO-LUMO excitation by light.

In realistic photodevices, the coupling of the energy levels to the baths is not totally asymmetric. In this case, whether the photodevice operates as a solar cell or photoconductor depends on asymmetry of the couplings and Coulomb interaction. Below, we analyse first a symmetric device, showing that it always operates as a photoconductor and highlighting the role of the Coulomb interaction $U$ in its operation. By tuning the asymmetry of the junction, we then study the solar cell regime in the presence of $U$.

\subsection{Photoconducting regime}

Henceforth, we consider that (i) the chemical potential of the fermionic baths lies between the HOMO and LUMO levels $\epsilon_{\mathrm{H}} \leq \mu \leq \epsilon_{\mathrm{L}}$ (see Fig.~\ref{fig1} (a)); (ii) a single coupling rate defines our device $\Gamma^{l}_{\mathrm{H}} = \Gamma^{r}_{\mathrm{L}} \equiv \Gamma$ and $\Gamma^{l}_{\mathrm{L}} = \Gamma^{r}_{\mathrm{H}} = z~\Gamma$, where $0 \leq z \leq 1$ is a symmetry parameter with $z=1,0$ corresponding to a symmetric or totally asymmetric device, respectively. While in previous works \cite{Rutten2009} the system was fully asymmetric (corresponding to $z=0$), the role of asymmetry is critical in photodevices  \cite{Galperin2012}, and we thus allow here a general framework which can continuously cross from the fully symmetric to the fully on-symmetric cases.

In Fig.~\ref{fig1} (b) we show the conductance $G \equiv dJ/dV |_{V=0}$ of the symmetric photodevice as a function of $U$ for different values of the photon rate $\nu$. For a symmetric device, we always find that it works as a photoconductor $\mathcal{Q} < 0$. While in the dark the device has negligible conductance regardless of $U$, in the presence of light we see an exponential increase in conductance around the transport gap $U \simeq \mu - \epsilon_{\mathrm{H}}$. Such a mechanism for photoconductance has been reported in symmetric molecular junctions and can be understood as a shift of the energy levels towards the chemical potential caused by exciton binding (see Fig.~\ref{fig1} (a) and Ref.~\cite{Dubi2017}). Its impact on thermodynamic quantities is shown in Fig.~\ref{fig2} (a)--(c) for finite $V > 0$, where we plot the entropy production rate in Eq.~\eqref{entropyproduction}, the absolute value of the coefficient of performance in Eq.~\eqref{performance} and the signal-to-noise ratio (or precision) of the particle current $\mathrm{SNR} = J^2 / D$ where $D$ is the noise quantifying particle current fluctuations. The noise $D$ can be computed using the quantum master equation \eqref{masterequation}, as outlined in Ref.~\cite{Landi2024} (see also Appendix~\ref{app:snr}). We always find that the entropy production rate decreases with $U$ and increases with the photon rate $\nu$. However, the most relevant feature is the emergence of a plateau centered around the transport gap $U \in [\mu - \epsilon_{\mathrm{H}} - V/2,\mu - \epsilon_{\mathrm{H}} + V/2]$. The performance and SNR are maximal inside the plateau but close to zero outside it, while the entropy production is always finite. This means that although light is always a source of dissipation, it only triggers conductance inside the plateau. This validates Eq.~\eqref{performance} in capturing photoconductance and shows that Coulomb interaction is crucial for an increased performance and precision of a photoconductor.

\subsection{Solar cell regime}

\begin{figure}[t!]
    \centering
    \includegraphics[width=0.48\textwidth]{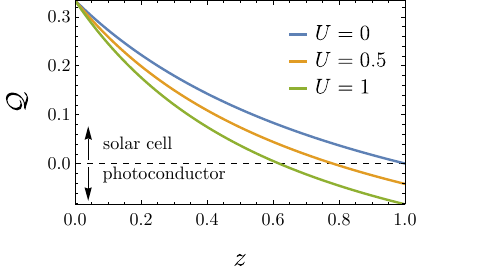}
    \caption{Coefficient of performance in Eq.~\eqref{performance} as a function of the symmetry parameters $z$ for different values of $U$, computed for $\nu = 100$. All other parameters are the same as in Figs.~\ref{fig1} and \ref{fig2}.}
    \label{fig3}
\end{figure}

While a symmetric device ($z = 1$) operates as a photoconductor $\mathcal{Q} < 0$, we have shown from Eq.~\eqref{performancerate} that a totally asymmetric device ($z = 0$) always operates as a solar cell at maximum efficiency. We illustrate in Fig.~\ref{fig3}, where we can see that the efficiency at $z=0$. Beyond the totally asymmetric case, i.e. in the regime $0 < z < 1$, we see that a solar cell decreases in efficiency and becomes sensitive to the presence of Coulomb interaction. By increasing $z$ beyond a certain threshold (which we could not determine analytically) the solar cell can be transformed into a photoconductor. Remarkably, we see that for certain fixed values of $z$ the Coulomb interaction can transform a solar cell into a photoconductor and vice-versa.

In Fig.~\eqref{fig2} (d)--(f) we show the entropy production rate, performance (efficiency) and SNR as a function of $U$ for a solar cell with a small degree of symmetry. The entropy production rate has a qualitatively similar behaviour as for the photoconductor shown before. However, inside the plateau around the transport gap $U \in [\mu - \epsilon_{\mathrm{H}} - V/2,\mu - \epsilon_{\mathrm{H}} + V/2]$ we see a decrease in efficiency and SNR -- the same mechanism responsible for photoconductance (which generates a current flowing with the potential bias) is detrimental for the operation of a solar cell. Interestingly, the SNR has a non-symmetric behaviour outside the plateau -- for small $U$ we observe a higher SNR than for large $U$. This shows that the Coulomb interaction energy can be beneficial in increasing the precision of a solar cell which is not totally asymmetric. 

\section{Conclusion}

We have introduced a model for a photodevice which can operate either as a solar cell or photoconductor. By studying its thermodynamics, we have shown that device performance can be quantified by a single coefficient in Eq.~\eqref{performance}. Both the asymmetry of the junction and Coulomb interaction energy play a crucial role in determining the operation regime and thermodynamic performance. Regarding the experimental relevance of our results, we showed that photoconductance based on exciton binding -- a mechanism first reported in symmetric molecular junctions \cite{Dubi2017} -- is associated with an increased thermodynamic performance and precision. Moreover, our results indicate that even asymmetric photoconductors -- as comprised e.g. of asymmetric molecular junctions -- can benefit in efficiency by high exciton binding energies. By going beyond the idealized fully asymmetric solar cell, we showed that the same mechanism responsible for photoconductance is detrimental for a solar cell. This last observation is in line with what is known for solar cells, particularly those composed of organic materials where a high exciton binding energy is detrimental for its efficiency \cite{Giebink2011,Nayak2019,Zhu2021}. However, our model makes no assumption regarding the nature of the junction which can be an organic molecule or inorganic quantum dots. Finally, our analysis focused on the fundamental mechanisms of operation -- an exploration of strong-coupling effects within the mesoscopic leads formalism \cite{Lacerda2023,Lacerda2023a,Bettmann2024}, including dissipation effect due to phonons and level broadening to the baths are left for future work. Overall, we expect that our minimal model contributes as a guide to design thermodynamically efficient photodevices.

\section*{Acknowledgements}

S.L.J. acknowledges the financial support from a Marie Skłodowska-Curie Fellowship (Grant No. 101103884). This work was supported by the EPSRC-SFI joint project QuamNESS and by SFI under the Frontier For the Future Program. J.G. is supported by a SFI Royal Society University Research Fellowship. 

\appendix

\section{Quantum master equation}
\label{app:qme}

We provide here some details on the quantum master equation presented in Eq.~\eqref{masterequation} of the main text. Since the quantum master equation is simply additive for multiple baths (see Eq.~\eqref{masterequation}), we focus our discussion on a system interacting with a single bath in full generality.

\subsection{Derivation \label{app:qmederivation}}

We consider a Hamiltonian $H = H_S + H_B + V_{SB}$ where $H_S$ and $H_B$ are the system and bath Hamiltonians, while $V_{SB}$ is the interaction between them. The aim is to obtain a differential equation for the system of the form $d \rho_S(t)/dt = \mathcal{L}[\rho_S(t)]$, where $\rho_S(t) = \mathrm{Tr}_B[\rho(t)]$ is the density operator of the system obtained by tracing out the state of the composite system $\rho(t)$ whose evolution is dictated by the unitary operator $U(t)=e^{-i H t/\hbar}$. In order to derive the aforementioned equation, we assume (i) weak-coupling (Born approximation), implying that the unitary evolution is truncated to second-order in the interaction $V_{SB}$; (ii) Markov approximation, implying that the bath has fast correlation times compared to the system's relaxation time; (iii) secular approximation, implying that the natural timescale of the system (dictated by the inverse frequencies associated with $H_S$) is much shorter than the system's relaxation time \cite{CohenTannoudji1998,Breuer2007,Milburn2008,Rivas2012}. 

We first decompose the interaction into
\begin{align}
    V_{SB} = \sum_{\alpha} A_{\alpha} \otimes B_{\alpha} = \sum_{\alpha, \omega} A_{\alpha}(\omega) \otimes B_{\alpha}
\end{align}
where $A_{\alpha}$ and $B_{\alpha}$ are system and bath interaction (self-adjoint) operators and
\begin{align}
    A_{\alpha}(\omega) = \sum_{\epsilon'-\epsilon = \hbar\omega} \Pi(\epsilon) A_{\alpha} \Pi(\epsilon')
\end{align}
are eigenoperators of the system's Hamiltonian $[H_S, A_{\alpha}(\omega)] = -\hbar \omega A_{\alpha}(\omega)$ defined through the projectors $\Pi(\epsilon)$ onto the energy eigenspace with energy eigenvalue $\epsilon$. Under the aforementioned assumptions, the quantum master equation (in the Schr\"odinger picture) becomes \cite{CohenTannoudji1998,Breuer2007,Milburn2008,Rivas2012}
\begin{align}
    \label{app:generalqme}
    \frac{d \rho_S(t)}{dt} & = \mathcal{L}[\rho_S(t)] \nonumber \\ & = -\frac{i}{\hbar}[H_S + H_{LS}, \rho_S(t)] + \mathcal{D}[\rho_S(t)] \; ,
\end{align}
where the Lamb-Shift Hamiltonian commutes with the Hamiltonian of the system $[H_{LS},H_S]=0$, thus providing a renormalization of the system's energy; its contribution is thus ignored for the subsequent discussion. The crucial part is the dissipator $\mathcal{D}$ given by
\begin{align}
    \label{app:diss}
    \mathcal{D}[\cdot] = \sum_{\alpha,\beta,\omega} \gamma_{\alpha \beta}(\omega) \bigg[ A_{\beta}(\omega) \cdot A^{\dagger}_{\alpha}(\omega) - \frac{1}{2} \{ A^{\dagger}_{\alpha}(\omega) A_{\beta}(\omega), \cdot \} \bigg]  
\end{align}
and accounts for the irreversible evolution of the system induced by the bath. The positive matrix 
\begin{align}
    \gamma_{\alpha \beta}(\omega) = \frac{1}{\hbar^2} \int^{\infty}_{-\infty} dt~ e^{i \omega t}~\mathrm{Tr}_B[B^{\dagger}_{\alpha}(t) B_{\beta}(0) \rho_B]
\end{align}
is the Fourier transform of two-time correlation functions of the bath, where $O(t) = e^{i H_B t / \hbar} O e^{-i H_B t / \hbar}$ and the state of the bath is assumed to be stationary $[\rho_B, H_B] = 0$. If the stationary state is is the thermal state $\rho_B = e^{-\beta H_B} / Z_B$ with $Z_B = \mathrm{Tr}_B[e^{-\beta H_B}]$, it can be shown that $\rho_S = e^{-\beta H_S}/Z_S$ is a steady state of the quantum master equation \cite{Breuer2007,Rivas2012}. Note that in the presence of multiple baths in thermal equilibrium with different temperatures (as in the main body of our work), the steady state is no longer the thermal state.

\subsubsection*{Example: Two-level system interacting with radiation}

We can now apply the general master equation to describe a two-level system interacting with thermal light. The photon bath has Hamiltonian
\begin{align}
    H_B = \sum_{\Vec{k},\lambda=1,2} \hbar \omega_{k} b_{\lambda}^{\dagger}(\Vec{k}) b_{\lambda}(\Vec{k}) \; ,
\end{align}
where $\Vec{k}$ is the wavevector with norm $k = |\Vec{k}|$, $\lambda$ is the polarization and $b_{\lambda}(\Vec{k}), b^{\dagger}_{\lambda}(\Vec{k})$ are bosonic destruction and creation operation. The interaction is
\begin{align}
    V_{SB} = - \Vec{D} \cdot \Vec{E} \; ,
\end{align}
where the electric field operator is given by
\begin{align}
    \Vec{E} = i \sum_{\Vec{k},\lambda=1,2} \sqrt{\frac{2 \pi \hbar \omega_k}{V}} \Vec{e}_{\lambda}(\Vec{k}) [ b_{\lambda}(\Vec{k}) - b^{\dagger}_{\lambda}(\Vec{k}) ] \; ,
\end{align}
with $\Vec{e}_{\lambda}(\Vec{k})$ is the transverse unit polarization vector and $V$ is the quantization volume. Moreover, the two-level system is defined by the operators
\begin{align}
    H_S & = \epsilon_g \ket{g}\bra{g} + \epsilon_e \ket{e}\bra{e} \\
    \Vec{D} & = \Vec{d} \ket{g}\bra{e} + \Vec{d}^{*} \ket{e}\bra{g} \; .
\end{align}
In the last expression, $g, e$ denote the ground and excited states, while $\Vec{d} = \braket{g|\Vec{D}|e}$ is the matrix element of the dipole operator. Taking the continuous limit and carrying out the calculations \cite{CohenTannoudji1998,Breuer2007,Milburn2008,Rivas2012}, Eq.~\eqref{app:diss} becomes
\begin{align}
    \label{app:opticalqme}
    & \mathcal{D}[\cdot] = \nu[n(\epsilon_e - \epsilon_g) + 1] \bigg[ \ket{g}\bra{e} \cdot \ket{g}\bra{e}^{\dagger} - \frac{1}{2} \{ \ket{g}\bra{e}^{\dagger} \ket{g}\bra{e}, \cdot \} \bigg] \nonumber \\
    & + \nu n(\epsilon_e - \epsilon_g) \bigg[ \ket{e}\bra{g} \cdot \ket{e}\bra{g}^{\dagger} - \frac{1}{2} \{ \ket{e}\bra{g}^{\dagger} \ket{e}\bra{g}, \cdot \} \bigg] \; .
\end{align}
In the last expression $n(u) = [\exp  (\beta u) - 1]^{-1}$ is Bose-Einstein distribution (since we assumed the radiation to be in the thermal state) and the photon rate $\nu$ is defined by
\begin{align}
    \label{app:photonrate}
    \nu = \frac{4 \omega^3 |\Vec{d}|^2}{3 \hbar c^3} \; ,
\end{align}
where $\omega = (\epsilon_e - \epsilon_g)/\hbar$ the natural frequency of the two-levels. From Eq.~\eqref{app:opticalqme}, we thus see that the number of photons absorbed per unit time is $\nu n(\epsilon_e - \epsilon_g)$ while the number of photons emitted per unit time (through spontaneous and stimulated emission) is $\nu [n(\epsilon_e - \epsilon_g) + 1]$ \cite{CohenTannoudji1998,Breuer2007,Milburn2008,Rivas2012}. Eq.~\eqref{app:opticalqme} can also be used phenomenologically to model e.g. molecular junctions interacting with light. In that case, the photon rate $\nu$ is a fitting parameter which can be inferred from the incoming photon flux and molecular cross section \cite{Dubi2017}.

\subsection{Jump operators}

We now focus on the photodevice model which is the focus of this work. An orthonormal basis for the Hilbert space of the system can be written in Fock space as $\{ \ket{0} , \ket{1} , \ket{2}, \ket{3} \}$, where $\ket{0}$ is the vacuum state, $ \ket{1} \equiv c^{\dagger}_{\mathrm{H}} \ket{0}$ is the state of one electron in HOMO, $ \ket{2} \equiv - c^{\dagger}_{\mathrm{L}} \ket{0}$ is the state of one electron in LUMO and $\ket{3} \equiv c^{\dagger}_{\mathrm{L}} c^{\dagger}_{\mathrm{H}} \ket{0}$ is the doubly occupied state. The Hamiltonian of the system is then simply $H_S = 0 \ket{0}\bra{0} + \epsilon_{\mathrm{H}} \ket{1}\bra{1} + \epsilon_{\mathrm{L}} \ket{2}\bra{2} + (\epsilon_{\mathrm{H}} + \epsilon_{\mathrm{L}} + U) \ket{3}\bra{3}$. Under the assumptions discussed in Sec.~\ref{app:qmederivation}, the dissipator of each bath becomes additive and is given by $\mathcal{D}_{\alpha}[\cdot] = \sum_k \big[ L^{\alpha}_{k} \cdot {L^{\alpha}_k}^{\dagger} - 1/2 \{ {L^{\alpha}_k}^{\dagger} L^{\alpha}_k,\cdot  \} \big]$ where $L^{\alpha}_k$ are jump operators belonging to bath $\alpha$. For each fermionic bath $\alpha = l,r$ we have 8 jump operators 
\begin{align}
    L^{\alpha}_{10} & = \sqrt{\Gamma^{\alpha}_{\mathrm{H}} f_{\alpha}(\epsilon_{\mathrm{H}})} \ket{1}\bra{0} \; , \label{jumpL1}\\    
    L^{\alpha}_{01} & = \sqrt{\Gamma^{\alpha}_{\mathrm{H}} [1-f_{\alpha}(\epsilon_{\mathrm{H}})]} \ket{0}\bra{1} \; ,\\
    L^{\alpha}_{32} & = \sqrt{\Gamma^{\alpha}_{\mathrm{H}} f_{\alpha}(\epsilon_{\mathrm{H}}+U)} \ket{3}\bra{2} \; ,\\
    L^{\alpha}_{23} & = \sqrt{\Gamma^{\alpha}_{\mathrm{H}} [1-f_{\alpha}(\epsilon_{\mathrm{H}}+U)]} \ket{2}\bra{3} \; ,\\
    L^{\alpha}_{20} & = \sqrt{\Gamma^{\alpha}_{\mathrm{L}} f_{\alpha}(\epsilon_{\mathrm{L}})} \ket{2}\bra{0} \; , \\    
    L^{\alpha}_{02} & = \sqrt{\Gamma^{\alpha}_{\mathrm{L}} [1-f_{\alpha}(\epsilon_{\mathrm{L}})]} \ket{0}\bra{2} \; , \\
    L^{\alpha}_{31} & = \sqrt{\Gamma^{\alpha}_{\mathrm{L}} f_{\alpha}(\epsilon_{\mathrm{L}}+U)} \ket{3}\bra{1} \; , \\
    L^{\alpha}_{13} & = \sqrt{\Gamma^{\alpha}_{\mathrm{L}} [1-f_{\alpha}(\epsilon_{\mathrm{L}}+U)]} \ket{1}\bra{3} \label{jumpL8} \; .
\end{align}
The quantity $\Gamma^{\alpha}_i$, with $i=\mathrm{H,L}$, are tunneling rates from the fermionic reservoirs and is given by \cite{Galperin2012,Cuevas2017}
\begin{align}
    \label{tunnelingrates}
    \Gamma^{\alpha}_i = \frac{2 \pi}{\hbar} \sum_{m_{\alpha}} |\lambda_{m_{\alpha}}|^2 \delta (\epsilon_{m_{\alpha}} - \hbar \omega) 
\end{align}
In the last expression, $\lambda_{m_{\alpha}}$ is the interaction strength of bath mode $m_{\alpha}$ with the system, $\epsilon_{m_{\alpha}}$ are the bath energies (see 
 Sec.~\ref{sec:photodevice}) and $\omega$ the system frequencies; the continuum limit of bath modes is usually taken. The rates in Eq.~\eqref{tunnelingrates} can then be inferred from the coupling strengths $\lambda_{m_{\alpha}}$, estimated e.g. through density functional theory or non-equilibrium Green's function methods \cite{Galperin2012,Cuevas2017}. For the photonic bath $\alpha = \gamma$, we have 2 jump operators
\begin{align}
    L^{\gamma}_{21} & = \sqrt{\nu n(\epsilon_{\mathrm{L}}-\epsilon_{\mathrm{H}})} \ket{2}\bra{1} \; ,\\    
    L^{\gamma}_{12} & = \sqrt{\nu [1+n(\epsilon_{\mathrm{L}}-\epsilon_{\mathrm{H}})]} \ket{1}\bra{2} \; .
\end{align}
The rate $\nu$ can be determined from the dipole strength as in Eq.~\ref{app:photonrate} or taken as a fitting parameter in phenomenological approaches.

\subsection{Rate equation}

The quantum master equation decouples populations from coherences in the eigenbasis of $H_S$. The latter decay and the former can be described in terms of a rate equation \cite{Breuer2007,Rivas2012}. The rate of change in populations of level $m \in \{ 0, 1, 2, 3 \}$ induced by the bath $\alpha$ is given by \begin{align}
    \label{ratepopulations}
    \dot{p}^{\alpha}_m(t) = \sum_{n} \braket{m|\mathcal{D}_{\alpha}[\ket{n}\bra{n}]|m} p_n(t) = R^{\alpha}_{mn} p_n(t) \; ,
\end{align}
where we omit the summation index and $p_n(t) \equiv \braket{n|\rho_S(t)|n}$ are the populations. Due to the Lindblad structure of the quantum master equation, the rate matrix $R^{\alpha}_{mn}$ has the form
\begin{align}
    R^{\alpha}_{mn} = r^{\alpha}_{mn} - \delta_{mn} \sum_i r^{\alpha}_{in} \; .
\end{align}
where $r^{\alpha}_{kl}$ are yrates associated with the jump operators of the bath $\alpha$ [e.g. $r_{10}^{\alpha} = \Gamma^{\alpha}_{\mathrm{H}} f_{\alpha}(\epsilon_{\mathrm{H}})$ for $\alpha = l,r$]. For $\alpha = l,r$ the rate matrix then reads
\begin{align}
R^{\alpha} = 
\begin{bmatrix}
-r^{\alpha}_{23} -r^{\alpha}_{13} & r^{\alpha}_{32} & r^{\alpha}_{31} & 0 \\
r^{\alpha}_{23} & -r^{\alpha}_{32} -r^{\alpha}_{02} & 0 & r^{\alpha}_{20} \\
r^{\alpha}_{13} & 0 & - r^{\alpha}_{31} - r^{\alpha}_{01} & r^{\alpha}_{10} \\
0 & r^{\alpha}_{02} & r^{\alpha}_{01} & - r^{\alpha}_{10} - r^{\alpha}_{20}
\end{bmatrix} \; ,
\end{align}
while for $\alpha = \gamma$ we have
\begin{align}
R^{\gamma} = 
\begin{bmatrix}
0 & 0 & 0 & 0 \\
0 & -r^{\gamma}_{12} & r^{\gamma}_{21} & 0 \\
0 & r^{\gamma}_{12} & - r^{\gamma}_{21} & 0 \\
0 & 0 & 0 & 0
\end{bmatrix} \; .
\end{align}
Note that $\sum_m R^{\alpha}_{mn} = 0$ for any fixed $n$ implies that the rate matrix is stochastic, which can be used to write Eq.~\eqref{ratepopulations} as follows
\begin{align}
    \label{currentpopulations}
    \dot{p}^{\alpha}_m(t) = \sum_{n \neq m} [r_{mn}^{\alpha} p_n(t) - r_{nm}^{\alpha}p_m(t)] = \sum_{n \neq m} j^{\alpha}_{mn}(t) \; ,
\end{align}
The last expression also defines a probability current which is anti-symmetric with respect to a given transition $j^{\alpha}_{mn}(t) = - j^{\alpha}_{nm}(t)$. We thus have the expressions for $\alpha = l,r$
\begin{align}
    \dot{p}^{\alpha}_0(t) & = j^{\alpha}_{01}(t) + j^{\alpha}_{02}(t) \; , \\
    \dot{p}^{\alpha}_1(t) & = j^{\alpha}_{10}(t) + j^{\alpha}_{13}(t) \; , \\
    \dot{p}^{\alpha}_2(t) & = j^{\alpha}_{20}(t) + j^{\alpha}_{23}(t) \; , \\
    \dot{p}^{\alpha}_3(t) & = j^{\alpha}_{32}(t) + j^{\alpha}_{31}(t) \; .
\end{align}
For the photonic bath $\alpha = \gamma$ we have 
\begin{align}
    \dot{p}^{\gamma}_1(t) & = j^{\gamma}_{12}(t) \; , \\
   \dot{p}^{\gamma}_2(t) & = j^{\gamma}_{21}(t) \; .
\end{align}
At steady state, the sum of the rates induced by all baths is zero $\sum_{\alpha} \dot{p}^{\alpha}_m = 0$ for all $m$ which gives rise to the following conservation laws for the currents
\begin{align}
    j^{l}_{01} + j^{l}_{02} + j^{r}_{01} + j^{r}_{02} & = 0 \label{conservationlaw1} \\
    j^{l}_{10} + j^{l}_{13} + j^{r}_{10} + j^{r}_{13} + j^{\gamma}_{12} & = 0 \label{conservationlaw2} \\
    j^{l}_{20} + j^{l}_{23} + j^{r}_{20} + j^{r}_{23} + j^{\gamma}_{21} & = 0 \label{conservationlaw3} \\
    j^{l}_{32} + j^{l}_{31} + j^{r}_{32} + j^{r}_{31} & = 0 \label{conservationlaw4} \; ,
\end{align}
where we drop the time-dependence when referring to steady state quantities.

\subsection{Thermodynamic quantities}

The rate matrix description allow us to compute expressions for the thermodynamic quantities. Using the particle number operator $N_S = 0 \ket{0}\bra{0} + \ket{1}\bra{1} + \ket{2}\bra{2} + 2\ket{3}\bra{3}$, we obtain the particle current associated to the fermionic baths $\alpha = l,r$ from Eq.~\eqref{particlecurrent} 
\begin{align}
    J^{\alpha}_N(t) & = \dot{p}^{\alpha}_1(t) +\dot{p}^{\alpha}_2(t) + 2 \dot{p}^{\alpha}_3(t) \nonumber \\
    & = j^{\alpha}_{10}(t) + j^{\alpha}_{20}(t) + j^{\alpha}_{31}(t) + j^{\alpha}_{32}(t) \label{particlecurrentrate}\; .
\end{align}
The photon heat current which can be computed from Eq.~\eqref{heatcurrent} (for $\mu_{\gamma} = 0$) is given by
\begin{align}
    J^{\gamma}_Q(t) & = \epsilon_{\mathrm{L}} \dot{p}^{\gamma}_2(t) + \epsilon_{\mathrm{H}}\dot{p}^{\gamma}_1(t) = (\epsilon_{\mathrm{L}} - \epsilon_{\mathrm{H}}) j^{\gamma}_{21}(t) \label{photonheatcurrentrate}\; .
\end{align}
A similar procedure yields the other heat currents which we do not show here explicitly.

\section{Coefficient of performance}
\label{app:performance}

We discuss here in more detail the coefficient of performance introduced in Eq.~\eqref{performance}. Using Eqs.~\eqref{particlecurrentrate} and \eqref{photonheatcurrentrate} at steady state, we can compute the coefficient of performance
\begin{align}
    \mathcal{Q} = \frac{V}{\epsilon_{\mathrm{L}} - \epsilon_{\mathrm{H}}} \Bigg[ \frac{j^{l}_{10} + j^{l}_{20} + j^{l}_{31} + j^{l}_{32}}{j^{\gamma}_{21}} \Bigg] \; .
\end{align}
Note that the prefactor is always positive in our model. Using Eq.~\eqref{conservationlaw2} to eliminate $j^{l}_{10}$ and then eliminating $j^{r}_{13}$ using Eq.~\eqref{conservationlaw4} we arrive at Eq.~\eqref{performancerate} of the main text.

\subsection{Solar cell}

From Eq.~\eqref{performancerate}, we see that our photodevice can only operate as a solar cell $\mathcal{Q} > 0$ if the condition
\begin{align}
    \label{solarcellcondition}
    j^{\gamma}_{21} > j^{r}_{\mathrm{H}} - j^{l}_{\mathrm{L}} \; 
\end{align}
is satisfied. Since $j^{r}_{\mathrm{H}}$ and $-j^{l}_{\mathrm{L}}$ represent currents flowing with the potential bias (into HOMO from bath $r$ and out of LUMO into bath $l$, respectively), the solar cell regime is only possible if the rate of HOMO-LUMO excitation by light exceeds the current flowing with the potential bias from the fermionic baths. Therefore, this current is always detrimental for the efficiency of a solar cell. 

However, this current vanishes for a totally asymmetric device where the fermionic bath $l$ is disconnected from HOMO and $r$ is disconnected from HOMO, i.e. $\Gamma^{l}_{\mathrm{L}} = \Gamma^{r}_{\mathrm{H}} = 0$. In this case we simply have $j^{\gamma}_{21} > 0$ (which is always guaranteed for a device that consumes light in order to operate) and the device always operates as a solar cell with maximum efficiency $0 <\mathcal{Q} = V/(\epsilon_{\mathrm{L}} - \epsilon_{\mathrm{H}}) \leq \eta_C$ as dictated by Eq.~\eqref{entropyproduction}. Such an efficiency is independent of the microscopic details of the photodevice. This is a signature of what is called thermodynamic strong coupling, where the particle current and (photon) heat current become proportional to each other \cite{Rutten2009,Broeck2013,denBroeck2015}. Indeed, for a totally asymmetric device $j^{l}_{20} = j^{l}_{31} = j^{r}_{10} = j^{r}_{32} = 0$ and Eq.~\eqref{particlecurrentrate} becomes $J^{l}_N = j^{l}_{10} + j^{l}_{32} = j^{\gamma}_{21}$, where the last equality follows from using the conservation laws in Eqs.~\eqref{conservationlaw1}--\eqref{conservationlaw4}. In summary, a totally asymmetric photodevice always operates as a solar cell at maximum efficiency $0 <\mathcal{Q} = V/(\epsilon_{\mathrm{L}} - \epsilon_{\mathrm{H}}) \leq \eta_C$ even in the presence of electron repulsion. As we have shown in the main text, beyond the ideal fully asymmetric device, electron repulsion plays an important role if the device is not fully asymmetric.

\subsection{Photoconductor}

From Eq.~\eqref{performancerate}, we see that our photodevice can only operate as a photoconductor $\mathcal{Q} < 0$ if the condition
\begin{align}
     j^{r}_{\mathrm{H}} - j^{l}_{\mathrm{L}} > j^{\gamma}_{21} \; 
\end{align}
is satisfied. If light is consumed at steady state $j^{\gamma}_{21} > 0$, we also see that the current flowing into HOMO from bath $r$ has to be larger than the current from out of LUMO into bath $l$. As we have shown in the main text, a totally symmetric device can only work as a photoconductor and its response is most pronounced when the electron repulsion energy matches the transport gap. This was reported experimentally in symmetric molecular junctions \cite{Dubi2017}.

\setcounter{figure}{3}

\begin{figure*}[t!]
    \centering
    \includegraphics{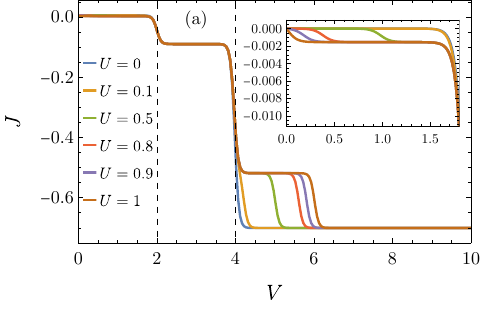}
    \includegraphics{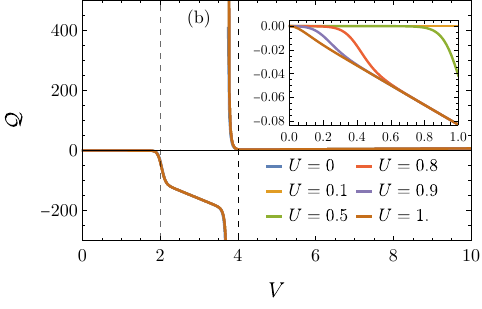}
        \includegraphics{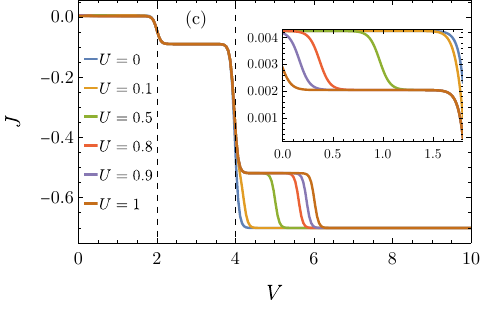}
    \includegraphics{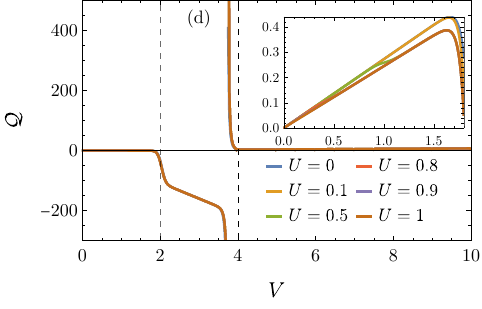}
    \caption{Steady state particle current $J$ and coefficient of performance $\mathcal{Q}$ as a function of bias $V$ for different values of $U$ and fixed photon rate $\nu$. Panels (a), (b) show a symmetric device ($z=1$) and (c), (d) show an almost asymmetric device ($z=0.1$). The dashed vertical lines separate different transport regimes: $0 \leq V \leq 2 |\epsilon_{\mathrm{H}}|$, $2 |\epsilon_{\mathrm{H}}| \leq V \leq 2 |\epsilon_{\mathrm{L}}|$ and $2 |\epsilon_{\mathrm{L}}| \leq V$. The insets show a zoom-in of the first regime chosen in the main text. The energies are $\mu = 0$, $\epsilon_{\mathrm{H}} = -1$ and $\epsilon_{\mathrm{L}} = 2$ and $\nu = 50$. All other parameters are the same as in Figs.~\ref{fig1} (b), \ref{fig2} and \ref{fig3}.}
    \label{fig:bias}
\end{figure*}

\section{Signal-to-noise ratio}
\label{app:snr}

The signal-to-noise ratio as defined in the main text $\mathrm{SNR} = J^2 / D$ involves computing the average particle current at steady state $J \equiv J_N^{\alpha}$ and its noise or diffusion coefficient $D \equiv D^{\alpha}$, where the latter captures particle current fluctuations at steady state. We describe briefly in this section how these fluctuations can be computed.

In general, one considers an open quantum system evolving with a quantum master equation of the form presented in Eq.~\eqref{masterequation} defined by the dissipators $\mathcal{D}_{\alpha}[\cdot] = \sum_k \big[ L^{\alpha}_{k} \cdot {L^{\alpha}_k}^{\dagger} - 1/2 \{ {L^{\alpha}_k}^{\dagger} L^{\alpha}_k,\cdot  \} \big]$. Fluctuations are studied by considering a counting variable $N^{\alpha}(t)$ which counts the number of events happening in the time interval $[0,t]$ associated to bath $\alpha$ (e.g. number of particles exchanged with bath $\alpha = l,r$ or photons emitted to bath $\alpha = \gamma$). The statistics of $N^{\alpha}(t)$ is determined by trajectory unravelling of the quantum master equation or through the formalism of full counting statistics \cite{Esposito2009,Schaller2014,Wiseman2014,Landi2024}. Following Ref.~\cite{Landi2024}, we can write the counting variable as a weighted sum
\begin{align}
    N^{\alpha}(t) = \sum_{k} w^{\alpha}_{k} N^{\alpha}_k(t) \; ,
\end{align}
where $N^{\alpha}_k(t)$ counts the events associated to channel $k$ in bath $\alpha$ in the interval $[0,t]$ and $w^{\alpha}_{k}$ are weights associated to each event. For instance, if we are interested in the number of particles exchanged with the fermionic baths $\alpha = l,r$ [see Eqs.~\eqref{jumpL1}--\eqref{jumpL8}] then $w^{\alpha}_{k} = 1$ for $k \in \{10, 32, 20, 31 \}$ (particles entering the system from the bath) and $w^{\alpha}_{k} = -1$ for $k \in \{01, 23, 02, 13 \}$ (particles leaving the system to the bath). The average current associated to $N^{\alpha}(t)$ is then
\begin{align}
    \label{currentcounting}
    I^{\alpha}(t) = \mathrm{E}\bigg[\frac{dN^{\alpha}(t)}{dt} \bigg] = \sum_{k} w^{\alpha}_k\mathrm{Tr}[L^{\alpha}_k \rho_S(t) L^{\alpha \dagger}_k]  \; ,
\end{align}
where $\mathrm{E}[X]$ denotes the expectation value of a random variable $X$ with respect to all possible trajectories or realizations. If we are interested in the number of particles exchanged with baths $\alpha = l,r$ as we discussed above, it is easy to see that $I^{\alpha}(t) = J^{\alpha}_N(t)$. On the other hand, the noise or diffusion coefficient associated with $N^{\alpha}(t)$ is defined by
\begin{align}
    \label{noisecounting}
    D^{\alpha}(t) = \frac{d}{dt} \mathrm{Var}[N^{\alpha}(t)] \; ,
\end{align}
where $\mathrm{Var}[X] = \mathrm{E}[X^2] - \mathrm{E}[X]^2$ is the variance and describes the fluctuations. Note that at steady state Eqs.~\eqref{currentcounting} and \eqref{noisecounting} become time-independent. In particular, the variance scales linearly with time $\mathrm{Var}[N^{\alpha}(t)] = D^{\alpha} t$ where $D^{\alpha}$ denotes the noise at steady state. As shown in Ref.~\cite{Landi2024} it can be computed through the compact expression
\begin{align}
    \label{noisefinal}
    D^{\alpha} = M^{\alpha} + 2 \int_{0}^{\infty} \Big\{ \mathrm{Tr}\big[\mathcal{J}^{\alpha} e^{\mathcal{L} \tau}\mathcal{J}^{\alpha}[\rho_S] \big] - I^{\alpha} \Big\}~d\tau \; .
\end{align}
In the last expression the quantity
\begin{align}
    M^{\alpha} = \sum_{k} (w^{\alpha}_k)^2\mathrm{Tr}[L^{\alpha}_k \rho_S(t) L^{\alpha \dagger}_k] \;
\end{align}
is called the dynamical activity and the superoperator
\begin{align}
    \mathcal{J}^{\alpha}[\cdot] = \sum_{k} w^{\alpha}_k L^{\alpha}_k \cdot L^{\alpha \dagger}_k \; ,
\end{align}
has already been used implicitly in Eq.~\eqref{currentcounting}. The Liouvillian superoperator $\mathcal{L}$ defines the quantum master equation in Eq.~\eqref{masterequation}. We computed the noise through Eq.~\eqref{noisefinal} for the case of the steady state particle current from the fermionic baths exemplified above. For an efficient numerical calculation, we made of use of vectorization as detailed in Ref.~\cite{Landi2024}. 

\section{Thermodynamic quantities vs. voltage bias}
\label{app:bias}

The data presented in Figs.~\ref{fig2} and \ref{fig3} was collected for a fixed bias $V \geq 0$ applied to the chemical potentials of each bath $\mu_l = \mu - V/2$ and $\mu_r = \mu + V/2$ with $\epsilon_{\mathrm{H}} \leq \mu \leq \epsilon_{\mathrm{L}}$. In this section, we illustrate the behaviour of thermodynamic quantities with the bias $V$. The particle current $J$ and coefficient of performance $\mathcal{Q}$ as a function of the bias are shown in Fig.~\ref{fig:bias} for the symmetric and almost asymmetric device ($z=1$ and $z=0.1$, respectively) and for a fixed photon rate. We always choose the chemical potential $\mu$ to be closer to HOMO than LUMO, so we can distinguish three operation regimes:

\begin{enumerate}
    \item Regime $\epsilon_{\mathrm{H}} \leq \mu_l \leq \mu_r \leq \epsilon_{\mathrm{L}}$ is the one analysed in the main text, corresponding to $0 \leq V \leq 2 |\epsilon_{\mathrm{H}}|$ in Fig.~\ref{fig:bias}. In this case, HOMO is mostly occupied while LUMO is mostly empty, so that transport can only emerge if an electron is excited to LUMO through photon absorption. In the symmetric device -- operating as a photoconductor $J, \mathcal{Q} < 0$ -- transport is only appreciable if $U$ is close to the transport gap. In the almost asymmetric device -- operating as a solar cell $J, \mathcal{Q} > 0$ -- light always triggers transport but $U$ decreases its performance. This regime necessarily implies $V \leq \epsilon_{\mathrm{L}} - \epsilon_{\mathrm{H}}$ so that a solar cell is always below ideal efficiency.
    \item Regime $\mu_l \leq \epsilon_{\mathrm{H}} \leq \mu_r \leq \epsilon_{\mathrm{L}}$ corresponds to $2 |\epsilon_{\mathrm{H}}| \leq V \leq 2 |\epsilon_{\mathrm{L}}|$. Transport can occur through HOMO, leading to an increase in the current flowing with the potential bias for both devices. The solar cell regime breaks down and both devices now operate as a photoconductor with similar behaviour. 
    \item Regime $\mu_l \leq \epsilon_{\mathrm{H}} \leq \epsilon_{\mathrm{L}} \leq \mu_r$ corresponds to $2 |\epsilon_{\mathrm{L}}| \leq V$ with transport occurring through both HOMO and LUMO, leading to a further increase in the current flowing with the potential bias. Note that the coefficient of performance diverges before changing its sign, which reflects the behaviour of the photon heat current $J_{Q}^{\gamma}$ (the latter is not shown in Fig.~\ref{fig:bias}). In this regime, an increase in $V$ results in a higher LUMO occupation so that photon emission starts to supersede absorption: thus $J_{Q}^{\gamma}$ changes continuously from positive to negative, leading to the observed divergence in $\mathcal{Q}$.
\end{enumerate}

\bibliography{references}

\end{document}